# Sputtering  based epitaxial growth and modeling of Cu/Si thin films

A.S. Bhattacharyya*, S. Kumar, S. Jana,  P.Y.Kommu, K.Gaurav, S.Prabha, V.S.Kujur, P.Bharadwaj

Centre for Nanotechnology, Central University of Jharkhand
Ranchi 835205, India

*Corresponding Author: 2006asb@gmail.com

Abstract—***Epitaxial copper thin films were deposited by magnetron sputtering. The adatoms during deposition are influenced by deposition parameters which cause variations in thin film properties. XRD and FESEM studies were done to get an insight into the growth mechanisms of the films. A modeling has been done on the epitaxial thin film growth with sputtering process. The parameters during sputtering like, sputtering yield, pressure, temperature, current density, deposition time were related and an attempt has been made to analyze the sputtering process.***

***Keywords- epitaxy; thin films; sputtering***

## INTRODUCTION

The transition metal copper in the form of thin film is applicable not only in electronic industry for interconnects and fabrication of devices but also useful as  a bioactive coating required for medical devices as copper is involved in various biological activity like embryonic development, mitochondrial respiration, regulation of hemoglobin  etc [1, 2]. Magnetron sputtering is an effective means of depositing nanocomposite as well as epitaxial thin films. The thin film adsorption process during magnetron sputtering is strongly influenced by the deposition parameters. The adatom mechanics can be tuned with the help of deposition parameters changing the thin film properties significantly. Copper thin films were also deposited by rf magnetron sputtering on Silicon substrates forming an epitaxy. These epitaxial films are highly favoured as conductor wires because they are free from grain boundary defects. They are also desirable for catalyst layer formation [3]. XRD and SEM studies provided an understanding of the growth mechanisms of the films. The Cu nuclei growth primarily took place on silicon by the Volmer-Weber mode which was evident from the 3D spherical island growth for the films. Deposition conditions like pressure, power, temperature and time have strong influence on the films. By variation of deposition pressure during sputtering, an alteration in the diffusional process occurring on the surface of the Cu thin films have been observed. This transport phenomenon occurring on the surface for copper (Cu) thin films deposited on silicon substrates leads to electro migration which is a major issue in electronic device manufacturing specially interconnects [4].

## EXPERIMENTAL

By variation of deposition pressure during sputtering, an alteration in the diffusional process occurring on the surface of the Cu thin films was observed. This transport phenomenon occurring on the surface for copper (Cu) thin films deposited on silicon substrates leads to electro migration which is a major issue in electronic device manufacturing specially interconnects [9]. Copper was deposited on Si (111) and Si (100) substrates by r.f. magnetron sputtering. The r.f. power was kept constant at 250 W with reflected power around 20 W at room temperature. The chamber pressure and deposition time were varied. The collisions taking place between the sputtered adatoms and the gas molecules in the chamber will depend upon the gas flow inside the chamber. The adatoms on reaching the substrate surface undergo diffusion to become thermodynamically and energetically stable. They also gain kinetic energy from the substrate temperature. We have discussed here both nanocomposite hard coatings  and epitaxial copper thin films deposited by magnetron sputtering and tried to throw some light on their growth mechanisms effecting their properties.

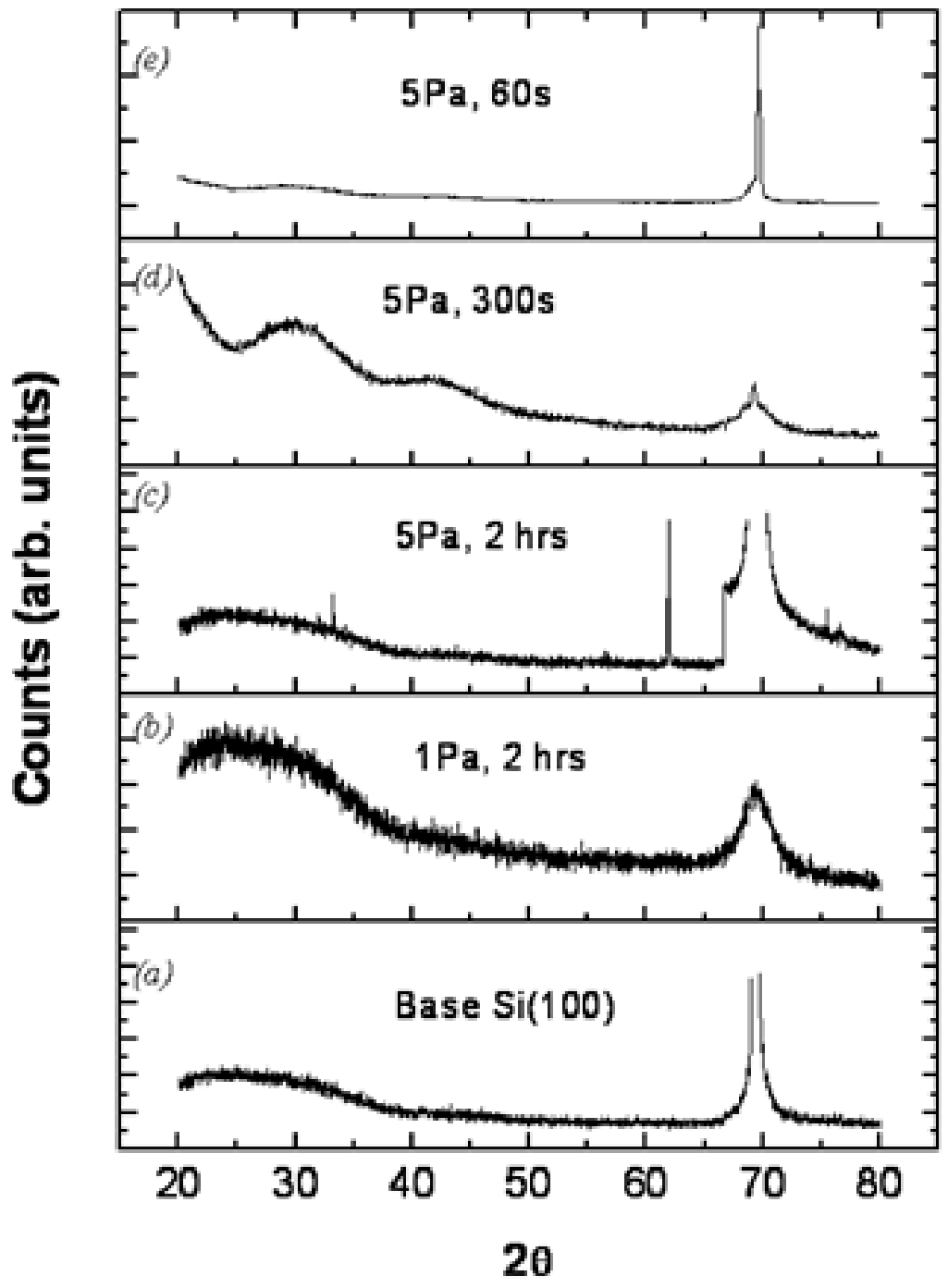

Figure 1. XRD of Cu/Si at different deposition conditions

*A. XRD studies*

XRD studies were performed on the thin films as well of the Si (100)) substrate [Fig 1]. The base Si (100) showed a characteristic peak at $68^o$. No major change was observed in the diffraction plot of Cu/Si (100) films which were deposited at 5 Pa pressure for 60s.The intensity of the peak coming from the base silicon was however much lower for films deposited for 300 s. An explanation for the lowering of the intensity could be the vibrations of the surface atoms about their equilibrium position and as a result the exact Bragg condition was not met. Another explanation could be the higher copper deposition reducing the substrate effect. The peaks due to copper were however visible only for films deposited at 5Pa pressure for 2 hrs. Changing the pressure to 1Pa while keeping the time of deposition again to 2hr interestingly showed absence of any major peak either from or Si showing chamber pressure playing a significant part in the surface properties of the film which was clear on performing the microscopic studies.

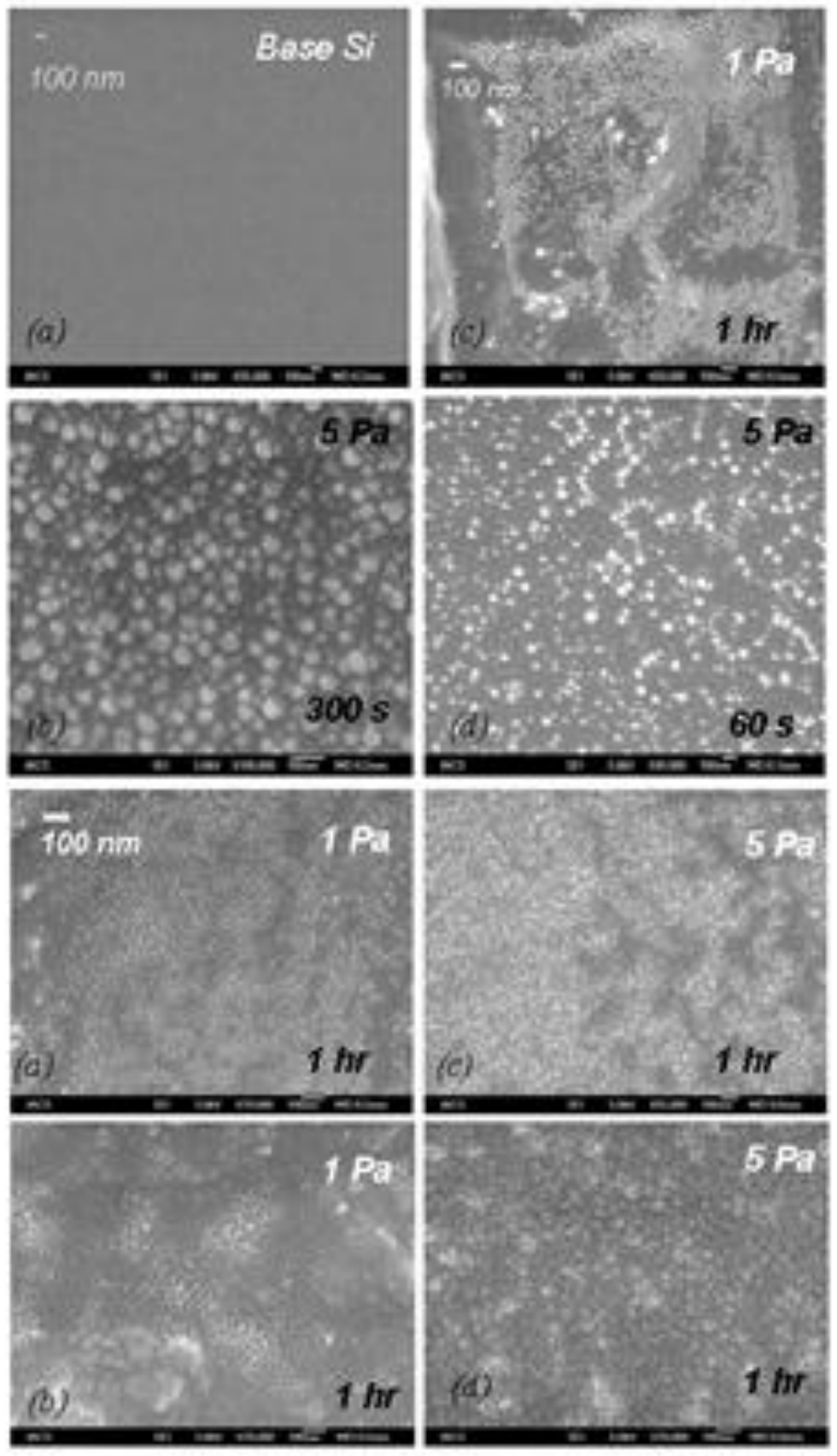

Figure 2. FESEM studies of base Si (100) and Cu/Si (100) deposited at different conditions

### *B. FESEM studies*

The FESEM studies for Cu/Si (100) are shown in Fig 3. The micrograph of the base silicon (100) surface is shown in Fig 2a. The deposition pressure and time are given on each micrograph. Fig 2c on the other hand shows the surface of Cu deposited on Si (100) for 1 hr at 1 Pa pressure at room temperature. The Cu nuclei primarily grew on silicon by the Volmer-Weber mode which was evident from the 3D spherical island growth for the films deposited for 60 s and 300s at 5 Pa pressure. This type of growth arises when the film atoms are more strongly bound to each other than to the substrate [5]. During deposition, the degree of crystallinity increased with deposition time and agglomeration and coalescence of grains took place. The irregular grain shape and orientation is attributed to the inhomogenous nature of sputtering process leading to larger size of the sputtered particles as well as the use of polycrystalline target which has an inhomogeneous microstructure. Single crystal targets however give homogenous films as the sputtered species have more uniform size and energy [3]. The substrate temperature in this case is also significant as the adatoms coming to the surface of the substrate acquire kinetic energy from the substrate temperature and may result in polycrystalline film. A higher substrate temperature may however lead to accelerated film oxidation [3]. Studies on the effect of Ar pressure on the microstructure and also on the formation of nanotwinned Cu films can be found in the literature [6, 7]. An abstract of the above discussion has been published [8].

## COMPUTATIONAL STUDIES

Sputtering is an effective means of depositing thin films. It based upon hitting a target (cathode) with energized ions to eject out adatoms from it. These adatoms get deposited on the substrate and form the film. The number of adatoms coming out per incident ion is called the sputtering yield. The sputtering yield is found to increase with ion energy which is quite obvious as more is the energy more will be the impact leading to more ejection of atoms from the target. An inert gas usually argon is introduced in an evacuated chamber and ionized. A plasma is formed in the process of ionization. The atomic density in plasma is usually of the order of $10^{-18}$ per $m^3$.

The sputtering rate i.e the amount of target materials sputtered per unit time is given by *zt = M/(rNAe) S jp* .where M is molar weight of the target [kg/mol]; r is the density of the material [kg/$m^3$]; NA is the 6.02 x $10^{26}$ 1/kmol; (Avogadro number); e is the 1.6x 10-19 As (electron charge)S is the sputtering yield (atom/ion) and jp: primary ion current density [A/$m^2$]. This rate divided by atomic diameter gives us the rate of sputtering in terms of atomic layers (AL) per second [9]. The adatoms sputtered from the target undergoes collision with the gas molecules was then multiplied with exp (-x/L) to get rough estimate of the deposited atomic layers per second. L(cm) is the mean free path of the sputtered atom which is related to the sputtering pressure P(Torr) and the molecular diameter Da of the sputtering gas (372 pm for Ar) $L = 2.303\times10^{-20}T/ (PDar^2)$ where T is the deposition temperature and x(cm) is the distance traversed by the adatom for deposition which was taken as 30 cm considering the target the substrate distance[10]. For copper (Cu) which has molar weight of 63.5 g/mol, density of 8.96 g/$cm^3$, sputtering yield of 1 for 1keV Ar ion. The radius of tungsten atom is 0.128 nm. The parameters which were varied to get a variation in the deposited ALs are given in Table 1.

Table 1: The parameters with their ranges which can be varied during the sputtering

| S. No | Parameter | Symbol | Range |
|---|---|---|---|
| 1. | Sputtering Yield | *S* | 0.5- 2.5 |
| 2. | Current density | *Jp* | 1-25 (mA) |
| 3. | Target to substrate distance | *x* | 30 – 100 (cm) |
| 4. | Sputtering time | *t* | 1.0 -10.0 (s) |
| 5. | Sputtering Pressure | *P* | $10^{-2}$ – $10^{-6}$ Torr |
| 6. | Sputtering Temperature | *T* | 300 – 700 K |

Each parameter was varied independently keeping the other parameters with a constant to see the effect on the deposition. The deposition pressure P was kept at $10^{-2}$ Torr, deposition temperature at 300K (RT) and deposition time *t* was for 30s. The variation of Atomic Layer depositions with sputtering yield at different deposition time is shown in Fig 3. The plasma current density in this case was kept fixed at 1 mA/$cm^2$. It can be observed that the rate of sputtering increases with increase in deposition time.

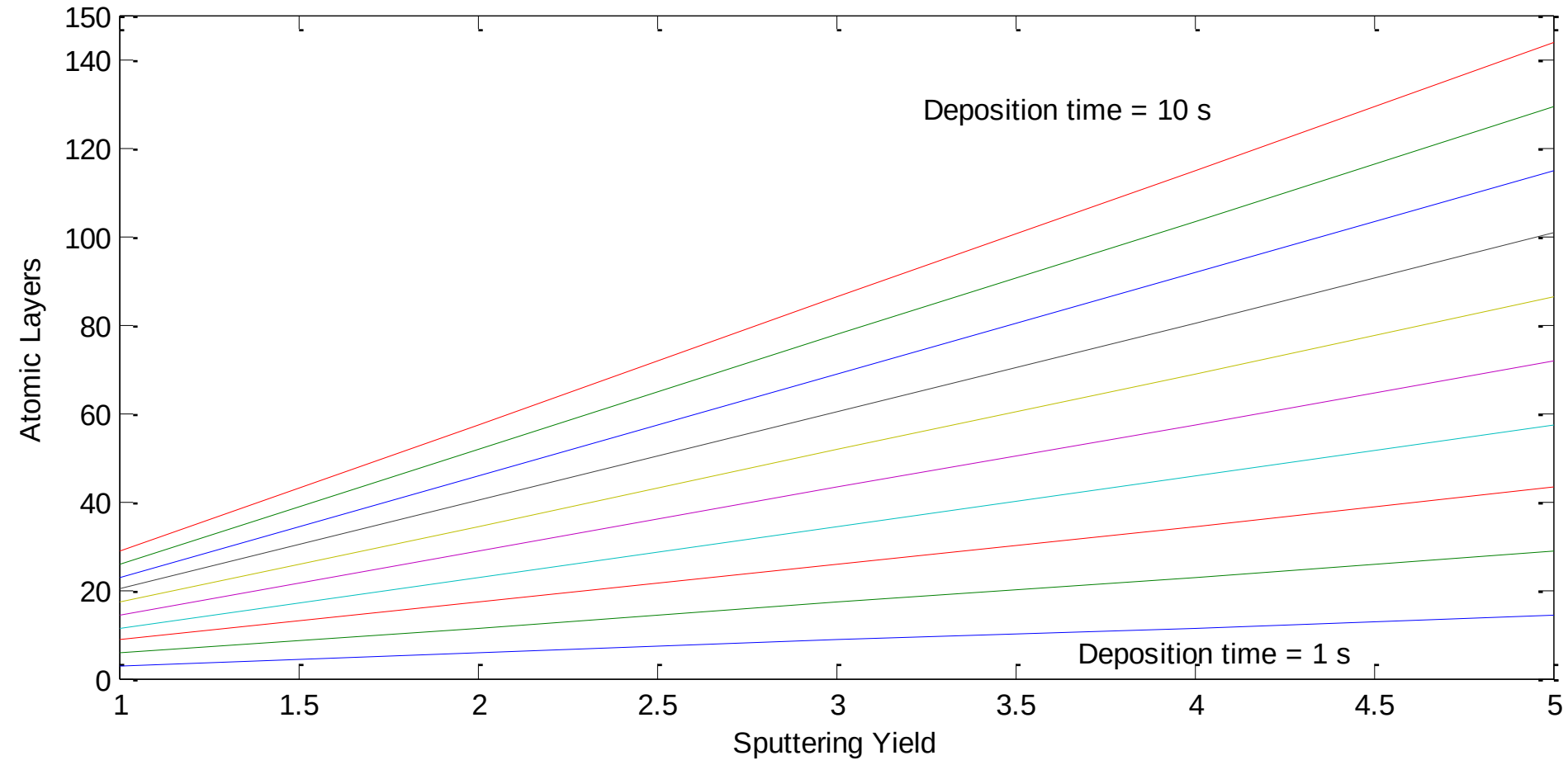

Fig 3: Atomic layers deposited with sputtering yield at different deposition times

The variation of Atomic layers deposited with source to substrate distance is shown in fig 4(a). The plot follows the exponential function. We have used a pressure of 1e-3 Torr. For different pressures however different plots are obtained as shown in fig 4(b). It shows that for lower and higher pressure the atomic layer deposition attains a constant value and is partially independent of the source to substrate distance. For pressures in the range of 1e-3and 1e-4, the variation is more pronounced. The explanation behind this result is the collisions the adatoms encounter during deposition. At low pressure or high vacuum, the mean free path of the adatoms is very large and thus the adatoms undergo very less collisions and so variation of target to substarte distance is insignificant. At higher pressure and lower vacuum conditions the mean free path being small, the adatoms undergo severe collisions which also makes the source to substrate distance insignificant. The Atomic layers decreased steadily with increase in temperature which can be attributed to the stress arising in the films due to increase in temperature causing lower adhesion with the substrate. In practical situation the variation is not so meticulous as lot of other factor like diffusion, coalescence of adatoms, thickness of films comes into the picture which requires further computation modeling. The increase in atomic layers deposited with current density is obvious from the fact that an increase current density usually leads to higher sputter rates.

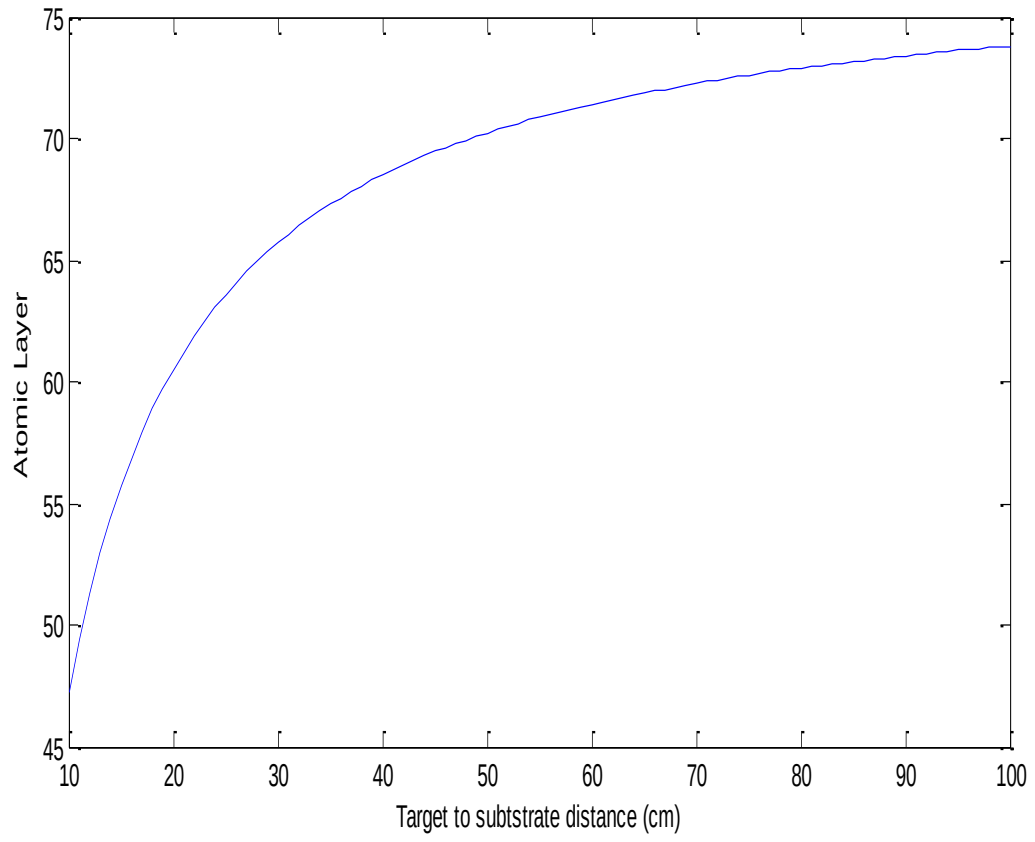

(a)

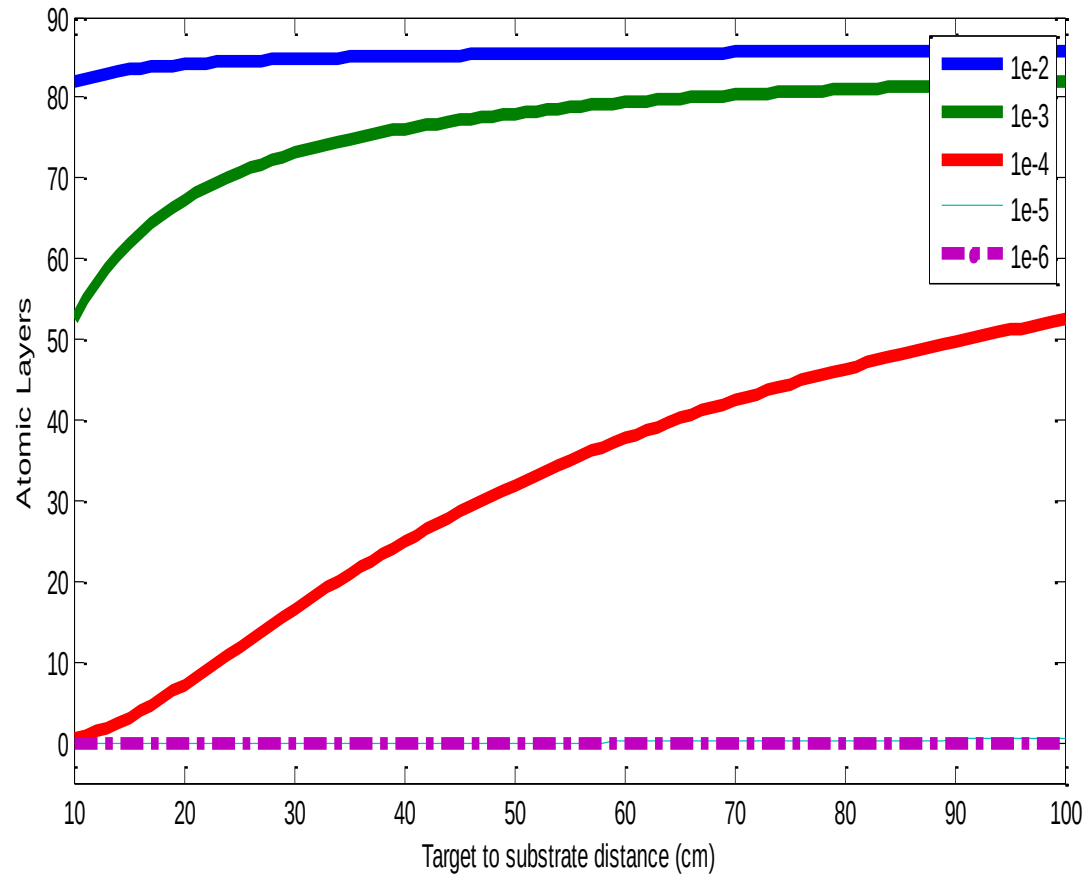

(b)

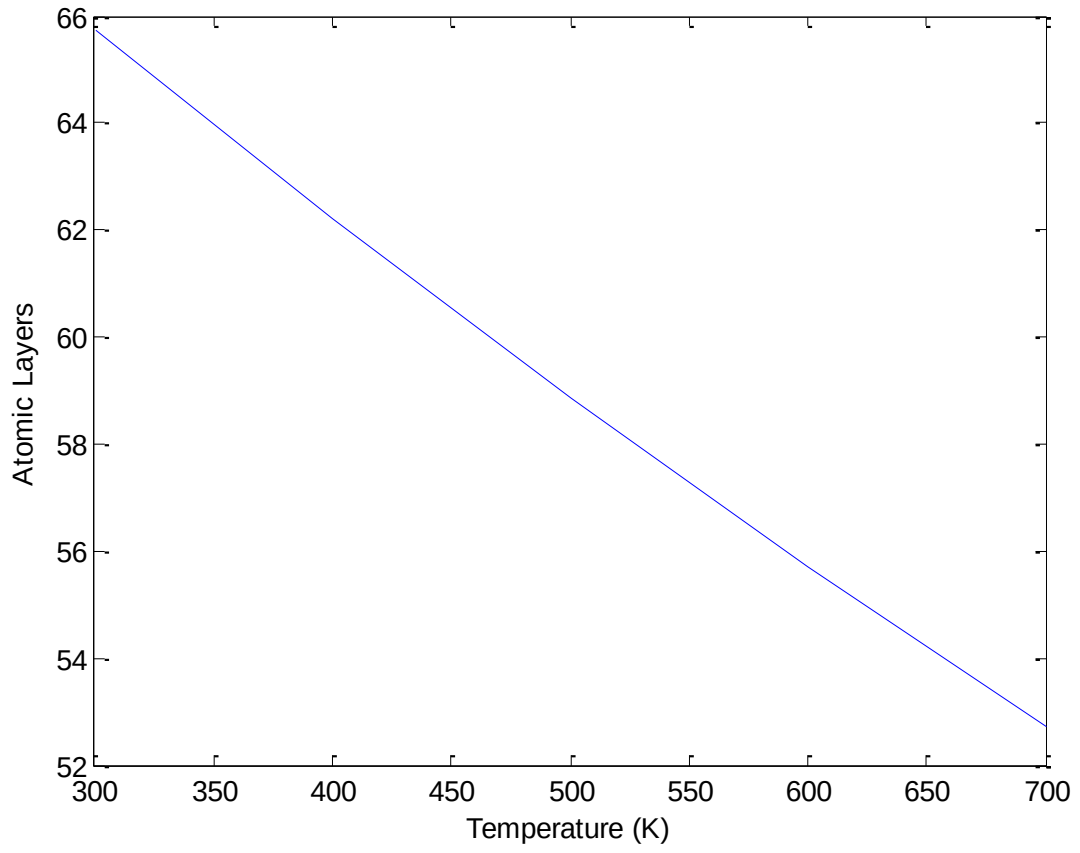

(c)

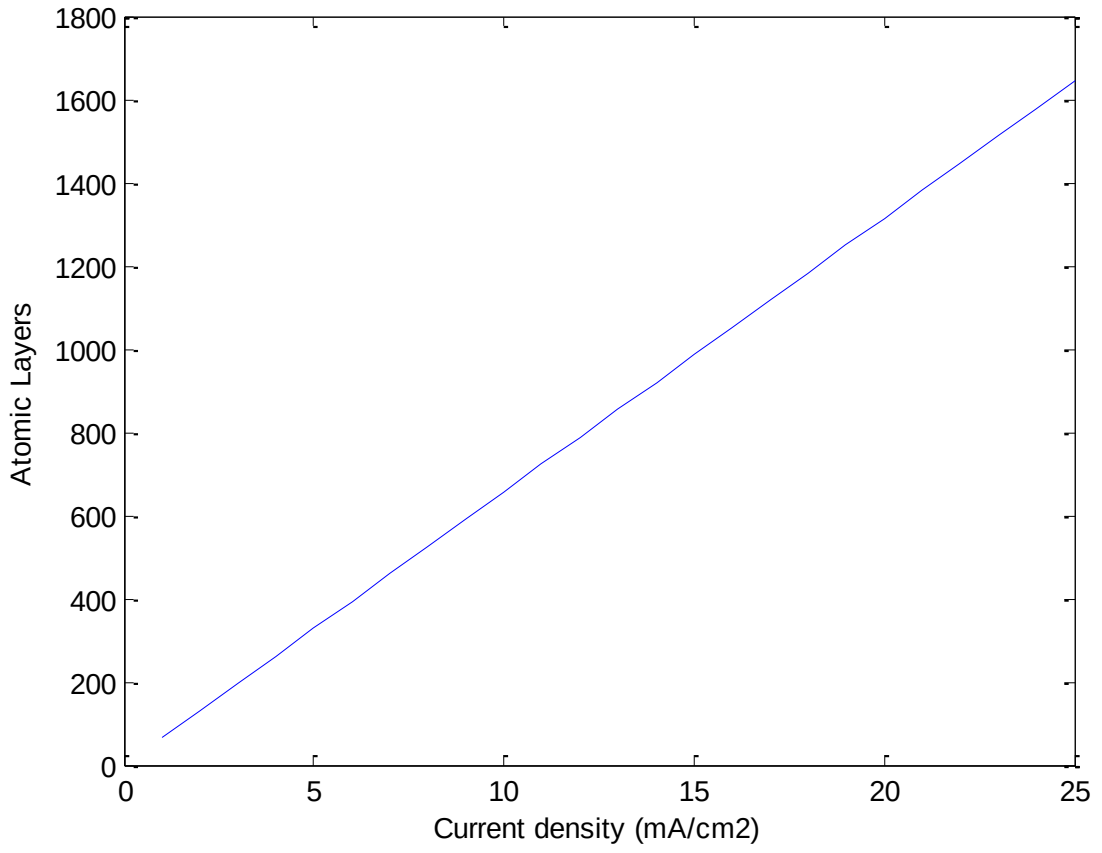

(d)

Fig 4: Atomic Layers deposited vs. (a) target to substrate distance (b) vs. target to substrate distance at different pressures Fig : Atomic Layers deposited vs. (c) temperature and (d) current density

ACKNOWLEDGMENT

The authors would like to thank Dr. B.N. Mondal and Mr. Chinmoy of IACS Kolkata for XRD and FESEM studies.